# Enhancement of optical properties and dielectric nature of $Sm^{3+}$doped $Na_2O$-ZnO-$TeO_2$Glass materials

Jyotindra Nath Mirdda[1], Subhadipta Mukhopadhyay[1], Kriti Ranjan Sahu[2*], Makhanlal Nanda Goswami[3]

[1]Dept. of Physics, Jadavpur University, Kolkata, 700032, India

[2]Dept. of Physics, Bhatter College, Dantan, Paschim Midnapore, West Bengal, 721426, India

[3]Dept. of Physics, Midnapore College, Paschim Midnapore, West Bengal, 721101, India

**Abstract**

Samarium doped $Na_2O$-ZnO-$TeO_2$ (NZT) glasses were prepared by the melt quenching method. The glass-forming ability and glass stability of prepared glass was estimated by Hruby parameter using Differential Thermal Analysis (DTA) and Thermo-gravimetric Analysis (TGA). The study of FTIR spectra and X-ray diffraction described the ionic nature and the amorphous pattern of glass respectively. The absorption peaks were observed for the transitions $^6H_{5/2}$-$^4P_{3/2}$ at 402 nm, $^6H_{5/2}$-$^4M_{19/2}$ at 418 nm, $^6H_{5/2}$-$^4I_{15/2}$ at 462 nm and $^6H_{5/2}$-$^4I_{11/2}$ at 478 nm in the absorption spectra. The optical band gap energy ($E_g$) was calculated and observed to be decreased from 2.95 eV to 1.50 eV with doping concentration. The visible emission band was observed in the $Sm^{3+}$ doped glass samples. The variation of dielectric constant with frequency was found to be independent for the frequency range 3 kHz - 2 MHz. The measurement of temperature-dependent dc conductivity showed Arrhenius type mechanism of conduction.

Corresponding Author: Kriti Ranjan Sahu, Email: kriti.basis2020@gmail.com, Mobile Phone: +91-8250808742

## 1. Introduction

Tellurium di-oxide$TeO_2$ is a promising glass network maker in the existence of alkali, alkaline earth and transition metal oxides (TMO) as modifiers, but it does not form glass itself. So, $TeO_2$ is familiar as a conditional glass producer, as it requires a modifier to produce the glassy state of the materials. Tellurite glass materials have some interesting properties like transparency at room temperature, hardness of satisfactory strength and attractive corrosion resistance [1-3]. Tellurite glasses are very interesting materials for linear and non-linear applications in optics, due to their significant characteristics such as low melting point, small phonon energy and large refractive index. It has also important for high dielectric constant, good chemical durability, high thermal stability, non-hygroscopic, with a large transmission window and the possibility to integrate a large quantity of rare-earth ions. It can be also applied as micro-lenses in photocopiers and mobile-phone cameras, IC photo masking in photolithography, photosensitive glass, hard disks, substrates for solar cells, artificial bones for mankind and dental transplants, etc. Glass materials are practically used for optical appliances due to their various compositions, physical isotropy and deficiency of grain boundaries. Tellurite glasses are useful to study by industrial researchers not only because of their methodological applications, but also owing to a fundamental significance in thoughtful their microscopic mechanisms [4, 5].

The luminescence properties of optical materials have been broadly observed not only for applications in phosphors, scintillators and laser crystals but also for scientific importance during the last two decays [6-12]. In presence of rare-earth ions, the glass materials are excellent luminescence intensity because of their emission power due to 4f-4f and 4f-5d electronic transitions [13-15].

Rare-Earth ions ($RE^{3+}$) doped tellurite glasses are applied due to their large potential in opto-electronic instruments such as solid-state lasers optical switches, broad-band amplifications, nonlinear optical devices, infra-red (NIR) laser windows, optical fibers three-dimensional displays, optical amplifiers, field emission displays, colour displays white light-emitting diodes, high-density optical data reading, biomedical diagnostics, microchip lasers, planar waveguides and high-density frequency domain optical data storage[16]. Samarium ($Sm^{3+}$) can be used as a dopant in glass hosts for powerful emissions in the visible area. Especially, the reddish-orange emission section from $Sm_2O_3$ doped NZT glasses holds strong photo-luminescence intensity, large emission cross-section and high quantum efficiency, which could be fit for laser applications. Therefore, $Sm^{3+}$ ions are the significant luminescent activators that are useful in illustrating the fluorescence properties because its $^4G_{5/2}$ level shows comparatively high quantum efficiency [4].

In this present work, the focus has been made to study properly the glass formation, structural, optical and electrical propertiesat room temperature of $Sm^{3+}$ doped $Na_2O$-$ZnO$-$TeO_2$ (NZT) glass compounds.

## 2. Experimental

The melt quenching process was used to prepare pure and $Sm^{3+}$doped $Na_2O$-$ZnO$-$TeO_2$ (NZT) glasses using research-grade initial composition Zinc Oxide (ZnO), Tellurium di-Oxide ($TeO_2$), Sodium Carbonate ($Na_2CO_3$) manufactured by Merck and Samarium Oxide ($Sm_2O_3$) made by LobaChemie. The mixing ratio of $Na_2O$, ZnO and $TeO_2$ is continued as 1:2:7 to make the host glass matrix. Samarium Oxide ($Sm_2O_3$) was added to the glass system as a dopant for (0-2) wt%. The standardized mixture of these precursors was obtained by grinding the ingredient powders in agate mortar. The mixture was put in an alumina crucible and the crucible was kept in an electrical furnace. The melt quenching method was reached by two stages of heating with the temperature at 400 °C for 1 hour and temperature at 475 °C for half an hour to prepare the good quality telluride glass. The cylindrical stainless-steel plate was used to hold the melted sample for quenching and the prepared glass material was placed again in the furnace at 400°C for 1hour to anneal the sample. The annealed glass was allowed to reach room temperature gradually through the slow cooling process to avoid thermal stress.

DTA/TGA of the initial mixtures (raw materials in the powder form) were carried out through the argon environment by using PerkinElmer Instrument (Pyris Diamond TG/DTA, thermo-gravimetric/differential thermal analyzer) for the temperature region of 30 °C to 650 °C with a scanning rate of 10°C/min. X-ray diffraction patterns were obtained using an X-ray diffractometer (RIGAKU model: Japan, XRD 6000, $\lambda$ = 1.5418 Å) with a slow-scanning rate 3°/min between the angle 10º and 70º for all the samples. FTIR spectrometer (HITACHI Model F-700) was used to detect the type of pure and doped glasses in the wave number region 400-3000 $cm^{-1}$. Absorption spectra and emission band of all the glasses were found using UV/VIS/NIR spectrophotometer (Perkin Elmer Lambda-35) for the wavelength range 400-800 nm and fluorescence spectrophotometer (HITACHI Model F-7000) for 300-600 nm at room temperature. The dielectric constants ($\varepsilon$) of prepared glass materials were measured using LCR-HiTESTER (HIOKI, Japan) for the wide frequency range of 100 Hz-8 MHz at room temperature. The temperature-dependent DC conductivity of glass samples was studied for the temperature range 36 °C-227 °C using constant voltage supply and current meter.

## 3. Results and Discussion

Figure1 shows the transparent $Sm_2O_3$ doped NZT glass samples. The colour of the pure NZT glass is white, while $Sm^{3+}$ doped NZT glasses are turned into yellowish colour due to the higher doping of $Sm^{3+}$ions in the NZT glass samples. There are no visible crystallites present in these transparent samples.

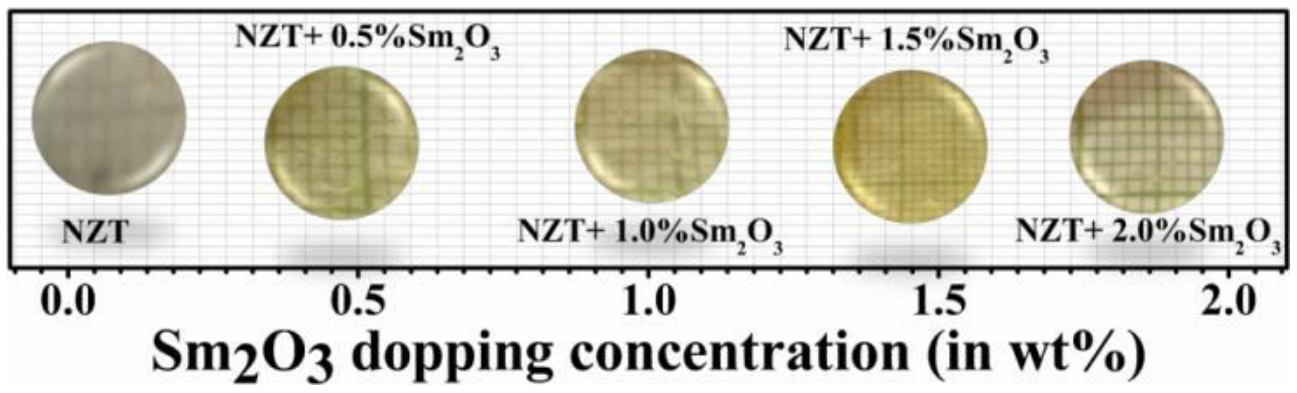

**Figure 1** Picture of $Sm_2O_3$ doped NZT glass sample.

### 3.1. Thermal properties

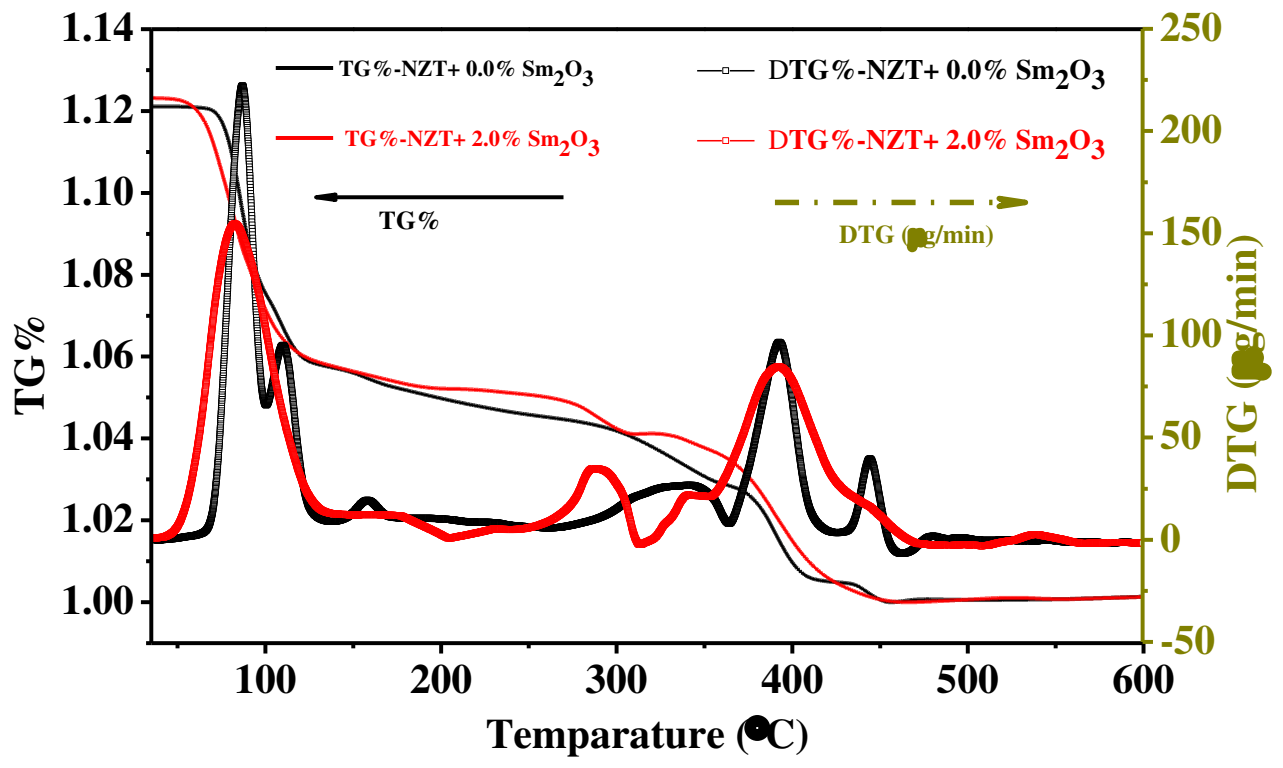

**Figure 2** DTA curves of $Sm_2O_3$ doped NZT glass samples.

Figure 2 shows the DTA curve of the pure and $Sm_2O_3$ doped (2.0% wt) NZT glasses from 40 °C to 550 °C at a heating rate of 10 °C/min. Three consecutive endothermic peaks have been found in DTA curve between (72 - 160) °C for pure and doped NZT glasses. These endothermic peaks on the DTA pattern are attributed to the desorption or exclusion of moisture. Generally endothermic peaks are evolved due to the melting point and exothermic peak indicates the crystallization of the melted glass [17]. Also, the exothermic peaks occurred when the $CO_2$ is released from the precursor powders and that has been found at around 406°C for pure and 416 °C for doped NZT glass.

**Table 1** Thermal parameters calculated from the DTA traces of pure and $Sm_2O_3$ doped NZT glasses.

| Sample Name | $T_g$ (°C) | $T_c$ (°C) | $T_m$ (°C) | $\Delta T = T_c - T_g$ | $H = (T_c-T_g)/(T_m-T_c)$ |
|---|---|---|---|---|---|
| NZT+0.0% $Sm_2O_3$ | 284 | 328 | 447 | 122 | 0.37 |
| NZT+2.0% $Sm_2O_3$ | 275 | 336 | 443 | 61 | 0.57 |

These curves also suggest that the phase transition occurred due to the transition of solid powder form to liquid form through melting at around 447 °C. The melting point of $TeO_2$ is 730°C which is decreased due to the presence of $Na_2CO_3$ mixed with ZnO [18]. The glass transition temperature ($T_g$) of pure and doped NZT glass materials has been observed to be 284°C and 275°. The decreased value of $T_g$ of the doped sample indicates onward formation of stable glass material. The thermal stability of the glasses is generally defined as a difference between glass transition temperature and crystallization temperature. Here, the estimated thermal stabilities ($\Delta T$) of the pure and $Sm^{3+}$ doped NZT glass materials have been tabulated in table 1. Glass thermal stability is stronger for a glass host to have $\Delta T$ as large as possible [17, 19]. The result specified here that the prepared $Sm^{3+}$ doped NZT glass holds good anti-crystallization ability and thermal stability. The glass-forming tendency can be explained using Eq. (1), where H is Hruby's parameter [20, 21].

$$H = \frac{Tc - Tg}{Tm - Tc}. \qquad (1)$$

The higher value of Hruby's parameter in the case of $Sm_2O_3$ doped NZT glass sample indicates the improvisation of glass-forming tendency due to doping of $Sm^{3+}$ ions.

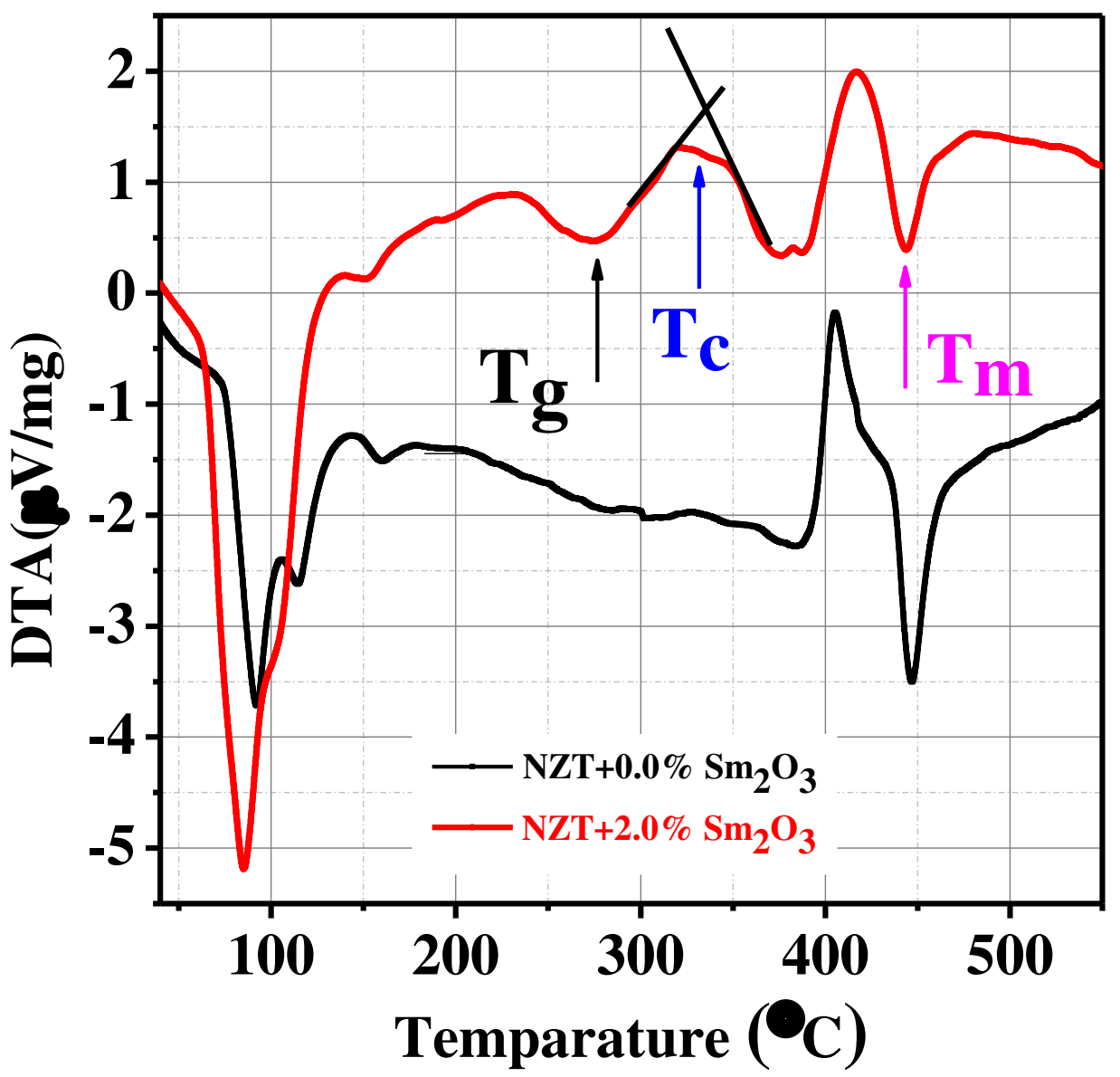

**Figure 3** TG with DTG curve of the pure and $Sm_2O_3$ doped NZT glasses.

Figure 3 shows the TG with DTG pattern of the pure and $Sm_2O_3$ doped NZT glasses. The weight loss of the precursor mixture has been observed for pure and doped glasses at different temperatures. TG/DTG curve illustrates the 5.37 % mass loss out of total mass due to exclusion of water, the release of moisture and other volatile substances in the temperature region 58-127 °C [22, 23]. The $CO_2$ is removed at the temperature range 380 °C to 406 °C for the decomposition of $Na_2CO_3$which is found from DTG curves for pure and doped samples. A little amount of mass is also lost within the temperature range of 435 °C to 450°C during the melting of the solid materials. No weight loss is detected beyond the temperature of 450 °C.

### 3.2. XRD

X-ray diffractogram of pure NZT and $Sm_2O_3$ doped glasses are shown in Fig. 4. It has been found from the figure that the XRD of all the samples demonstrate no discrete sharp peaks and a broad peak has been found in the region of the glancing angle at 24° to 34°. This broad peak in the XRD pattern indicates the non-crystalline materials as characteristics of typical long-range structural disorder of the samples. The broad continuous pattern of X-ray diffraction reveals the amorphous nature of these glasses [24, 25].

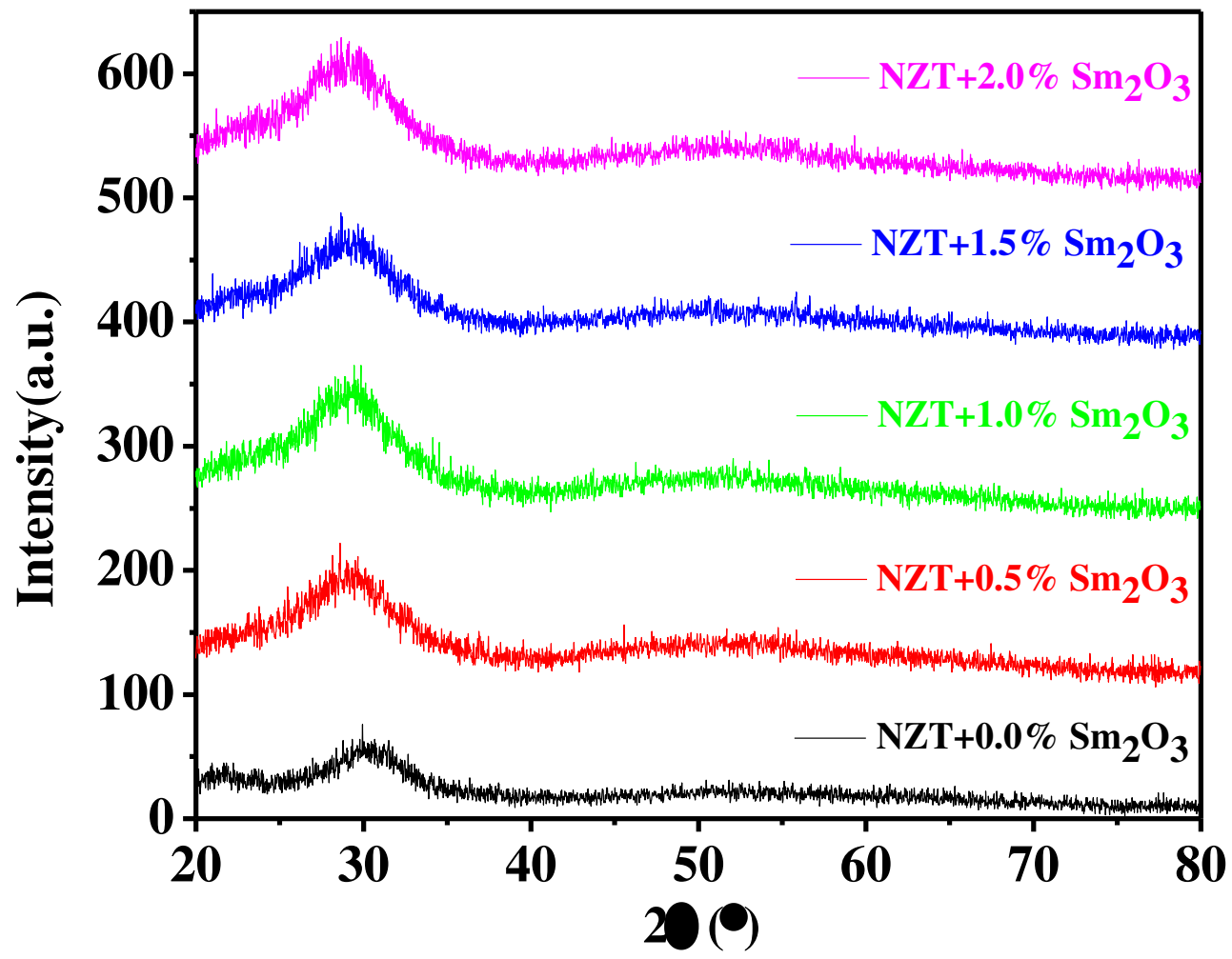

**Figure 4** XRD pattern of sample pure and $Sm^{3+}$ doped with NZT glass materials.

## 3.3. Optical Property

### 3.3.1. FTIR spectroscopy

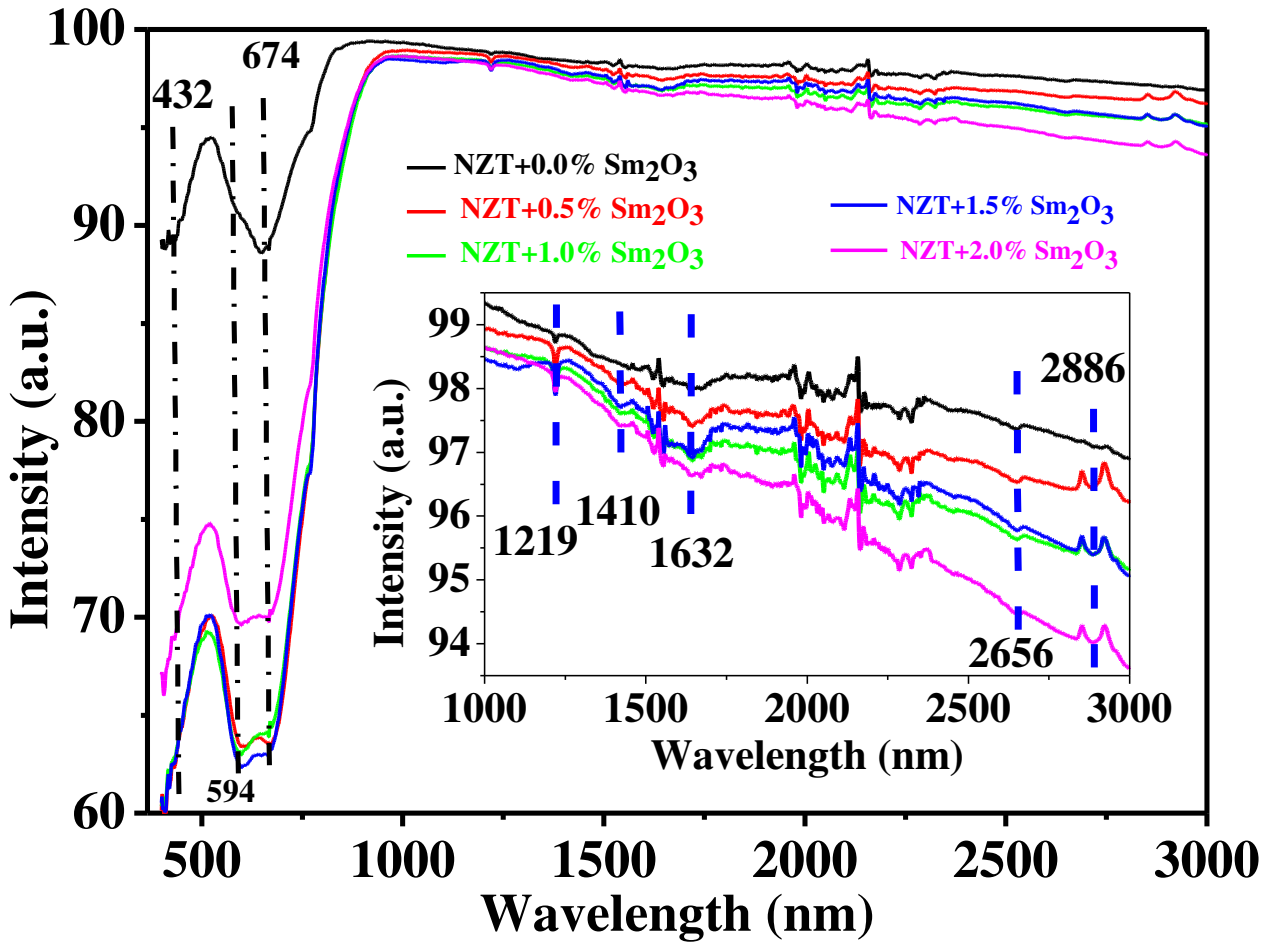

**Figure 5** FTIR spectroscopy of pure and various $Sm^{3+}$ doped with NZT glass materials.

Figure 5 represents the FTIR transmission spectra of pure and $Sm_2O_3$ doped NZT glasses in the wave number range 400-3000 $cm^{-1}$ at room temperature. The band at 424-440 $cm^{-1}$ comes out due to the symmetric stretching vibration of the Zn-O bond. $Zn^{2+}$ ions perform as an agent to split bonds and generate a modification in the glass structure. For this result, $TeO_{3+1}$ and $TeO_3$ can be twisted from $TeO_4$ units making non-bridging oxygen atoms [26-28]. This is also confirmed from the observation of glass transition temperature in DTA. The band is located at 570–601 $cm^{-1}$due to the stretching vibration of $TeO_4$ (trigonal pyramid) [29, 30]. It is observed that transmission intensity slightly changes due to the presence of $Sm_2O_3$ in the glass system. The band at 670–685 $cm^{-1}$ is observed due to the stretching vibration of $TeO_4$ trigonalbipyramid (tbp) [29]. The fundamental groups such as (- OH) bond, hydrogen bond and H-O-H bending appear in the higher region at 1632$cm^{-1}$[31]. The broad shoulder corresponds to the hydrogen bonds are observed around 2886 $cm^{-1}$.

**3.3.2. Absorption spectroscopy**

UV/Viz absorption spectra of pure and $Sm_2O_3$-doped NZT glasses are displayed in Fig. 6. The recording of absorption intensity has occurred in the visible range (350 - 540 nm) at room temperature with the band assignments. It is found from fig. 6, the absorption intensity increases gradually with the increase of doping concentration of $Sm_2O_3$. There is no transition occurred for the pure NZT glass sample whereas the absorption peaks are appeared at specific transitions in the case of $Sm^{3+}$ ions doping. The tail of the optical absorption band is not strongly identified due to the amorphous nature of glasses which is also established from XRD. The peak value of $Sm^{3+}$ doped NZT samples is obtained for the interaction of *4f -5d* electronic configurations [32]. It is observed that there are four absorption peaks for the transitions $^6H_{5/2}$-$^4P_{3/2}$ at 402 nm, $^6H_{5/2}$-

$^4M_{19/2}$ at 418 nm, $^6H_{5/2}$-$^4I_{15/2}$ at 462 nm and $^6H_{5/2}$-$^4I_{11/2}$ at 478 nm. Carnall et al. approved these transitions [33, 34].

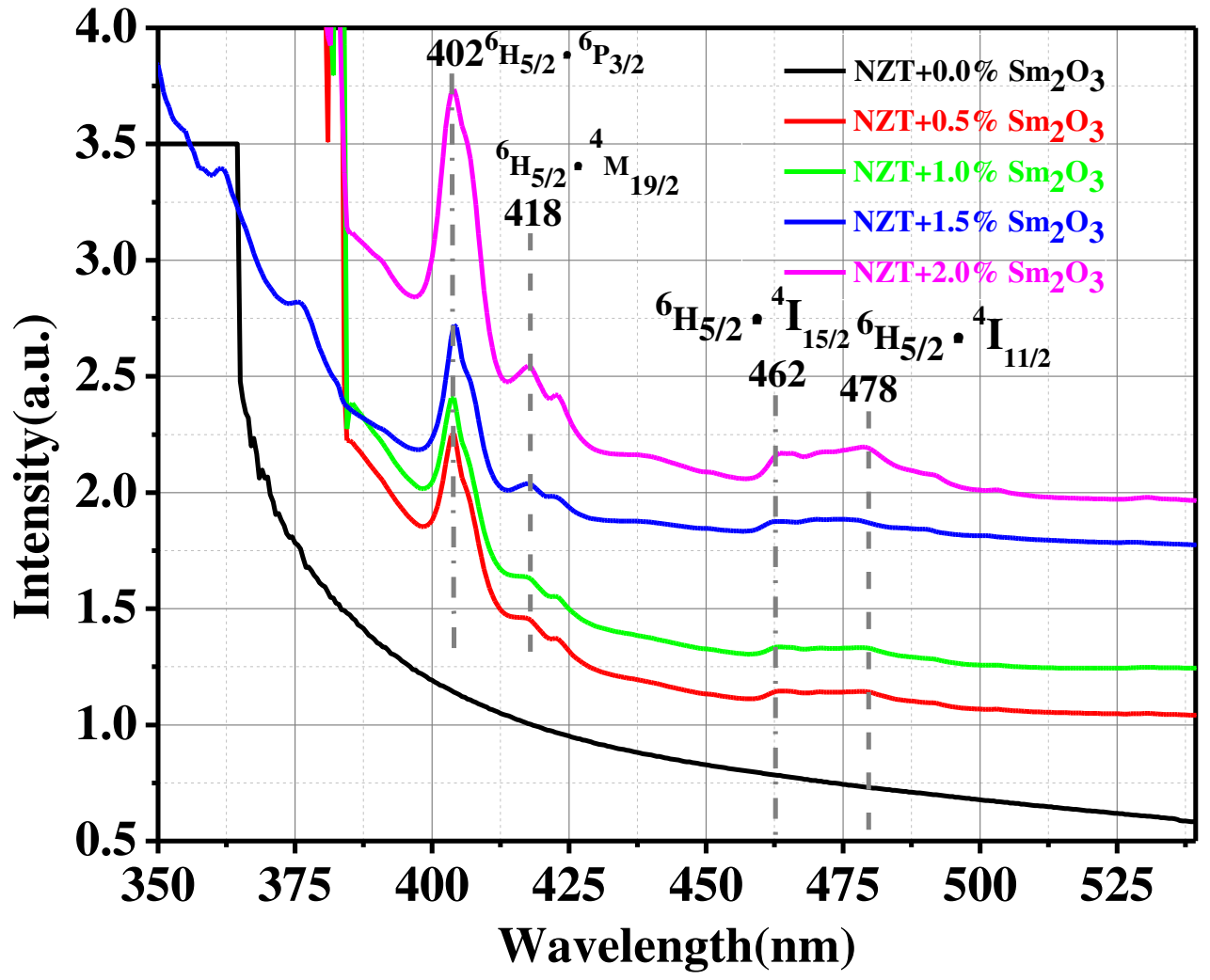

**Figure 6** Absorption spectroscopy of pure and $Sm^{3+}$ doped with NZT glass materials.

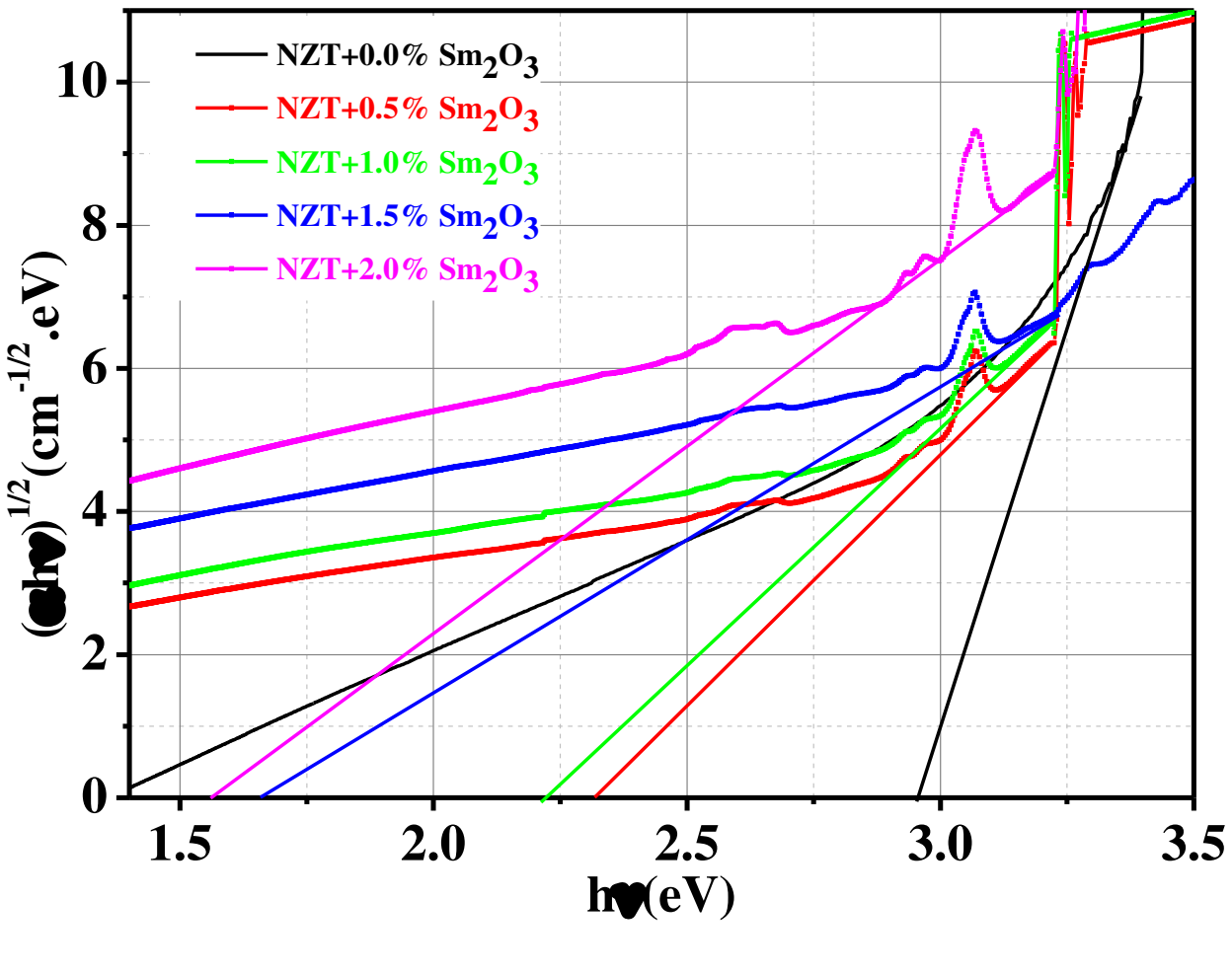

**Figure 7** Tauc’s plot for pure and $Sm^{3+}$ doped with NZT glass materials.

Figure 7 shows the variation of absorption coefficient α(ν) with corresponding energy. The absorption coefficient in many non-crystalline materials reflects the density of states at the band tails and can be determined from the optical absorption spectra using Eq. (2)

$$\alpha(\nu) = \left(\frac{A}{t}\right) \quad \dots \dots \dots \quad (2)$$

Where α(ν) is the absorption coefficient, A is the absorbance and t is the width of the sample

$$\alpha(\nu) = \frac{A\left(h\nu - E_g\right)^m}{h\nu} \quad \dots \dots \dots \quad (3)$$

**Table 2** Band gap energy, refractive index and other physical properties of pure and $Sm^{3+}$ doped NZT glasses.

| Samples Name | Band gap energy $E_g$ (eV) | Refractive index(n) | Inter Molecular distance ($r_i$) in Å | Polaron Distance ($r_p$) in Å | Field strength(F) × $10^{16}$ | Ion density ×$10^{20}$ |
|---|---|---|---|---|---|---|
| NZT+0.0% $Sm_2O_3$ | 2.95 | 2.41 | -- | -- | -- | -- |
| NZT+0.5% $Sm_2O_3$ | 2.28 | 2.63 | 14.98 | 6.037 | 4.125 | 2.97 |
| NZT+1.0% $Sm_2O_3$ | 2.22 | 2.65 | 13.86 | 5.586 | 4.819 | 3.76 |
| NZT+1.5% $Sm_2O_3$ | 2.12 | 2.69 | 19.51 | 7.862 | 2.432 | 1.35 |
| NZT+2.0% $Sm_2O_3$ | 1.58 | 2.95 | 20.41 | 8.228 | 2.220 | 1.18 |

Equation (3) gives the relation between absorption coefficient and photon energy. It has been determined the value of indirect band gap energy for m = 2 from the above equation (2) by plotting the absorption coefficient to zero absorption in the graph $(\alpha h\nu)^{1/2}$ vs hν. For the indirect transitions, the corresponding value of $E_g$ is obtained by extrapolating $(\alpha h\nu)^{1/2} = 0$. The values of $E_g$ for the glasses are tabulated in table 1. The decrease of $E_g$ with increasing $Sm^{3+}$ ions in the glass network can be explained by the creation of large number of NBO in the samples. The negative charges of oxygen ions in the NBO create active electrons with a higher value than that on the bridging oxygen (BO). These active electrons are distributed irregularly as because NBO's are softly bounded by the tellurite atoms [34-36].

The obtained values of the $E_g$ and the refractive index ($n$) can be estimated using Eq. (4) [36, 37].

$$\frac{n^2 - 1}{n^2 + 2} = 1 - \sqrt{\frac{E_g}{20}} \quad \dots \dots \dots \quad (4)$$

Refractive index increases with the increasing number of $Sm^{3+}$ ions due to the creation of non-bridging bonds. This effect can be explained by the dissociation of the bridging Te–O–Te bonds and created non-bridging bonds. The bond energies of non-bridging oxygen (NBO) bonds are lower with greater ionic character. The cation refractions have a higher magnitude for the NBO bonds are determined from this explanation [37].

### 3.3.3 Fluorescence Spectra

Figure 8 shows the fluoresce emission spectra for different concentrations of $Sm_2O_3$ doped with NZT glasses. The spectra have been recorded for samples with four different concentrations of $Sm^{3+}$ ions at an excitation wavelength 402 nm in the visible range of 500 - 700 nm at room temperature.

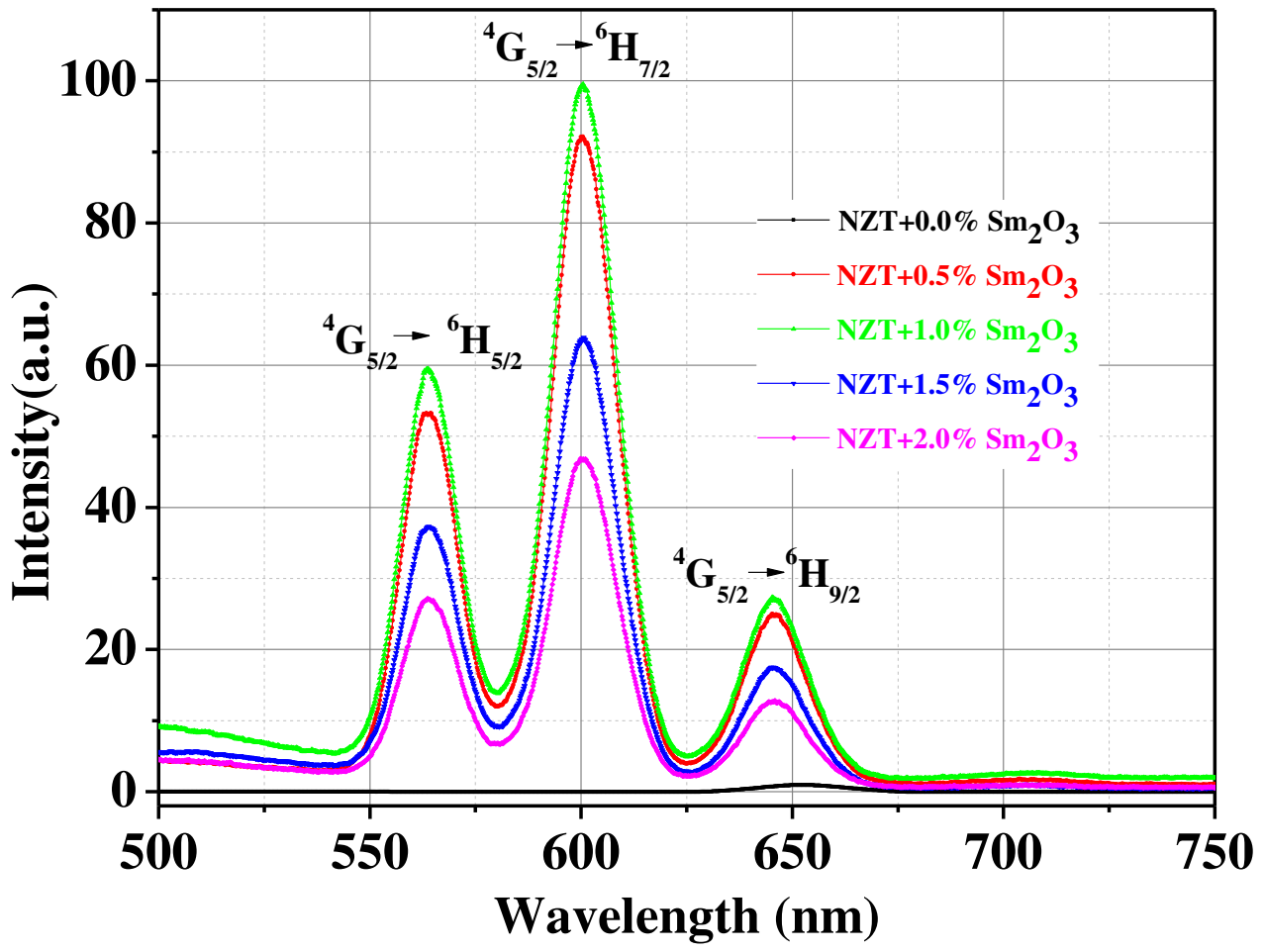

**Figure 8** Luminescence spectra of pure and $Sm^{3+}$ doped with NZT glass materials, excited at 402 nm.

Figure 8 displays the three peaks at the wavelengths 563 nm (green), 600 nm (orange) and 645 nm (red) with the matching to the transitions $^4G_{5/2}\rightarrow{}^6H_{5/2}$, $^4G_{5/2}\rightarrow{}^6H_{7/2}$, $^4G_{5/2}\rightarrow{}^6H_{9/2}$ of $Sm^{3+}$ ions respectively [16, 32, 38-40]. It is found that the emission band at 600 nm ($^4G_{5/2}\rightarrow{}^6H_{7/2}$) is the most outstanding band in all $Sm_2O_3$ doped NZT glasses.

The higher intensity of allowed transition $^4G_{5/2}\rightarrow{}^6H_{7/2}$ (600 nm) is occurred due to the contribution of magnetic dipole in the $Sm^{3+}$doped NZT glasses. In comparison to the intensity of the three emission peaks, the magnetic transition ($^4G_{5/2}\rightarrow{}^6H_{7/2}$) is greater than the intensity of electric dipole transitions ($^4G_{5/2}\rightarrow{}^6H_{5/2}$) and ($^4G_{5/2}\rightarrow{}^6H_{9/2}$) due to the ionic nature and symmetry of the system. The melt quenching of $Sm^{3+}$ doped NZT glass samples has also been confirmed from the emission band according to the increasing and decreasing intensity conforming to the quenching effect of various doping concentrations. Usually, the quenching effect is created from the non-radiative method consisting of multi-phonon relaxation and the non-radiative cross-relaxation may be obtained by two $Sm^{3+}$ ions through dipole-quadrupole or dipole-dipole or quadrupole-quadrupole interactions [41-45].

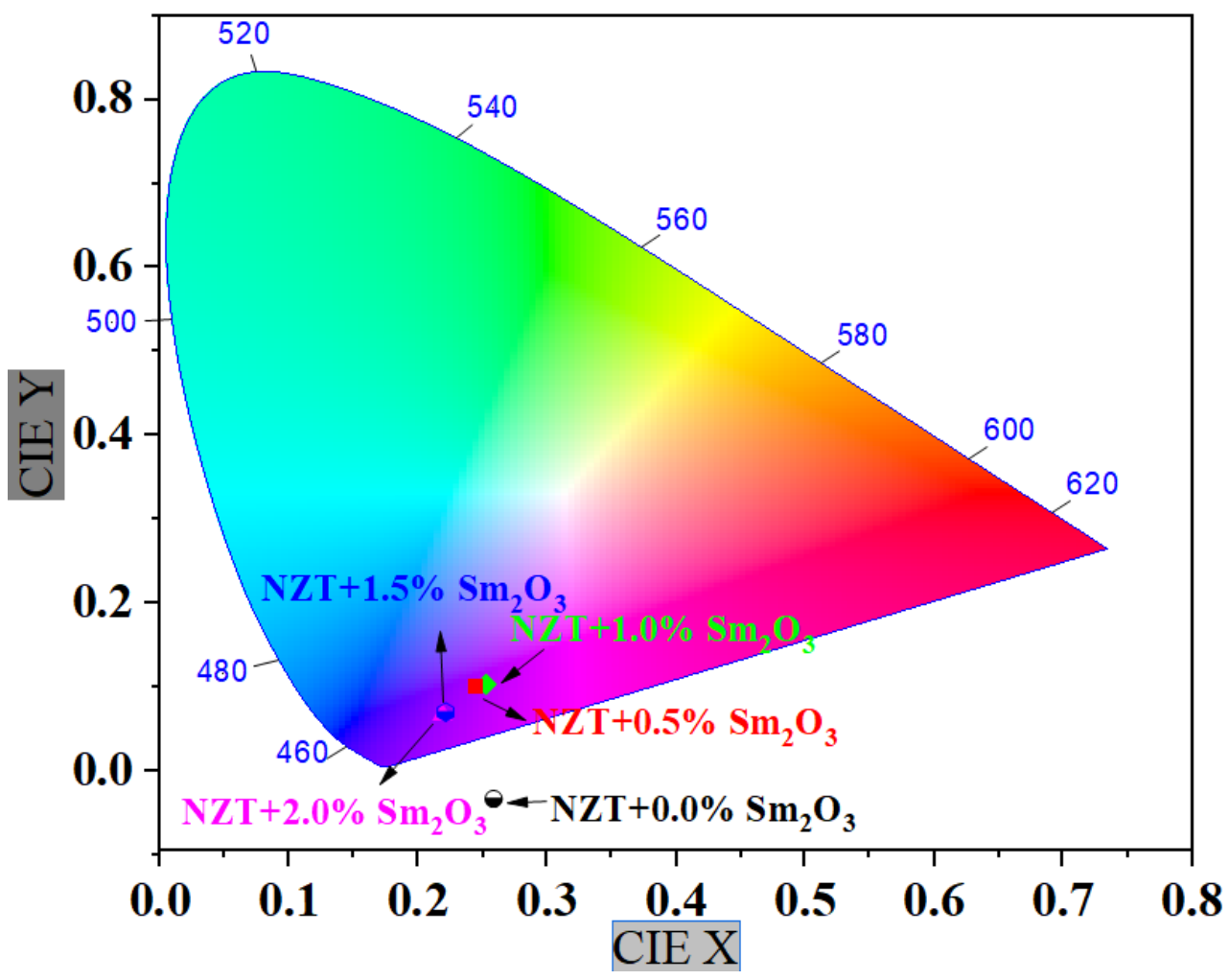

**Figure 9** CIE chromaticity color coordinate diagram of $Sm^{3+}$ doped with NZT glass materials, excited at 402 nm.

The scattering wavelengths of emission in the visible area can be quantified using CIE 1931 colour coordinates, which is a physiologically apparent colour vision of humans. The emission band of $Sm_2O_3$ doped NZT glass samples were plotted using CIE 1931 (Fig. 9). The obtained colour confirmed the different colours of emission of all materials. The colour coordinates was (0.25839, -0.03215), (0.2526, 0.1042), (0.21879, 0.06764), (0.24418, 0.1011), and (0.22123, 0.0711) for $Sm^{3+}$ doped (0.5% wt-2% wt) respectively. The non-emissive nature of the pure glass sample was confirmed from coordinates out the CIE plot. The changes in coordinates indicated that with increasing rare-earth oxide percent the emission shifted towards higher wavelengths.

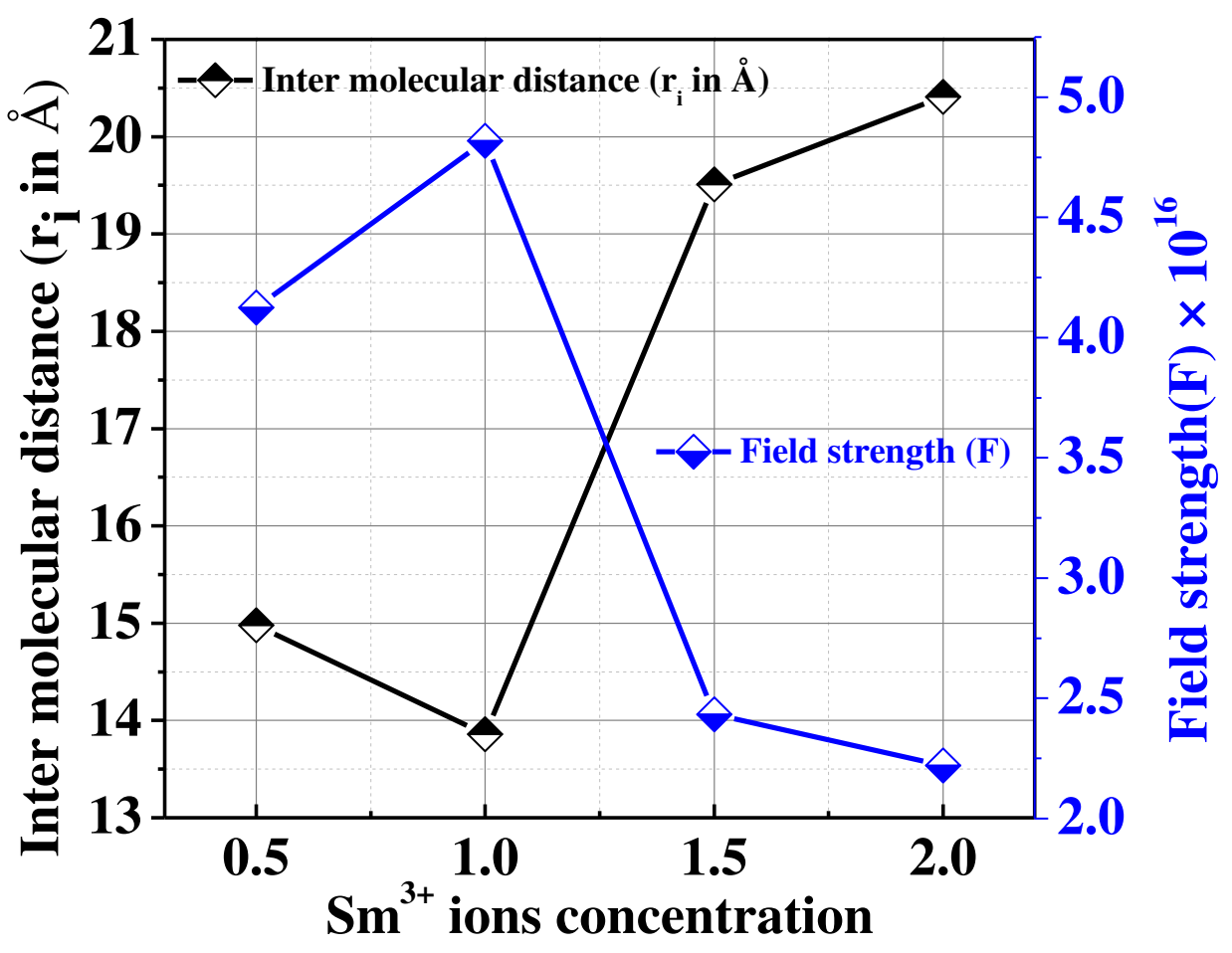

**Figure 10** Variation of intermolecular distance and field strength with $Sm^{3+}$ ions concentration.

$$r_i = \left(\frac{1}{N}\right)^{\frac{1}{3}} \quad \ldots \ldots \ldots \quad (5)$$

$$r_p = \left(\frac{\pi}{6N}\right)^{\frac{1}{3}} \quad \ldots \ldots \ldots \quad (6)$$

$$F = \left(\frac{Z}{r_p^2}\right) \quad \ldots \ldots \ldots \quad (7)$$

Figure 10 shows the variation of inter-ionic distance ($r_i$) and field strength ($F$) with ion concentration of $Sm^{3+}$ ions. Inter ionic distance ($r_i$) and polaron distance ($r_p$) decrease up to 1 wt% doping concentration of $Sm^{3+}$ ions. The field strength ($F$) between $Sm^{3+}$- $Sm^{3+}$ ions are increased with the ion concentration up to 1 wt% of $Sm^{3+}$ ions are obtained using Eq. (7). Again, the inter-ionic and polaron distance increase up to 2 wt% of $Sm^{3+}$ ions and correspondingly the field strength ($F$) decreases due to the variation of $Sm^{3+}$ ions which confirmed the fluoresce emission spectra. As the inter-ionic distance ($r_i$) decreases between two $Sm^{3+}$ ions for the compactness of glass structure, the strong interaction of $Sm^{3+}$ ions can transfer the excitation energy from one $Sm^{3+}$ ion to the other [32].

### 3.3.4. Cross Section

The absorption cross-section of the glass materials for the $^4G_{5/2}$–$^6H_{7/2}$ transition has been measured using Lambert-Beer formula

$$\sigma_{ab}(\lambda) = 2.303\frac{A}{Nt} \quad \ldots \ldots \ldots \quad (8)$$

Where A is absorbance, t is the width of the glass sample and $N$ is the density (ions/cm$^3$) of $Sm^{3+}$ ions of the telluride glass. Figure 11 shows the absorption cross-section is increasing for increasing the concentration of $Sm^{3+}$ ions in the NZT glass samples. The emission cross-section $\sigma_{emi}$ is calculated using MaCumber (1964) theory [46].

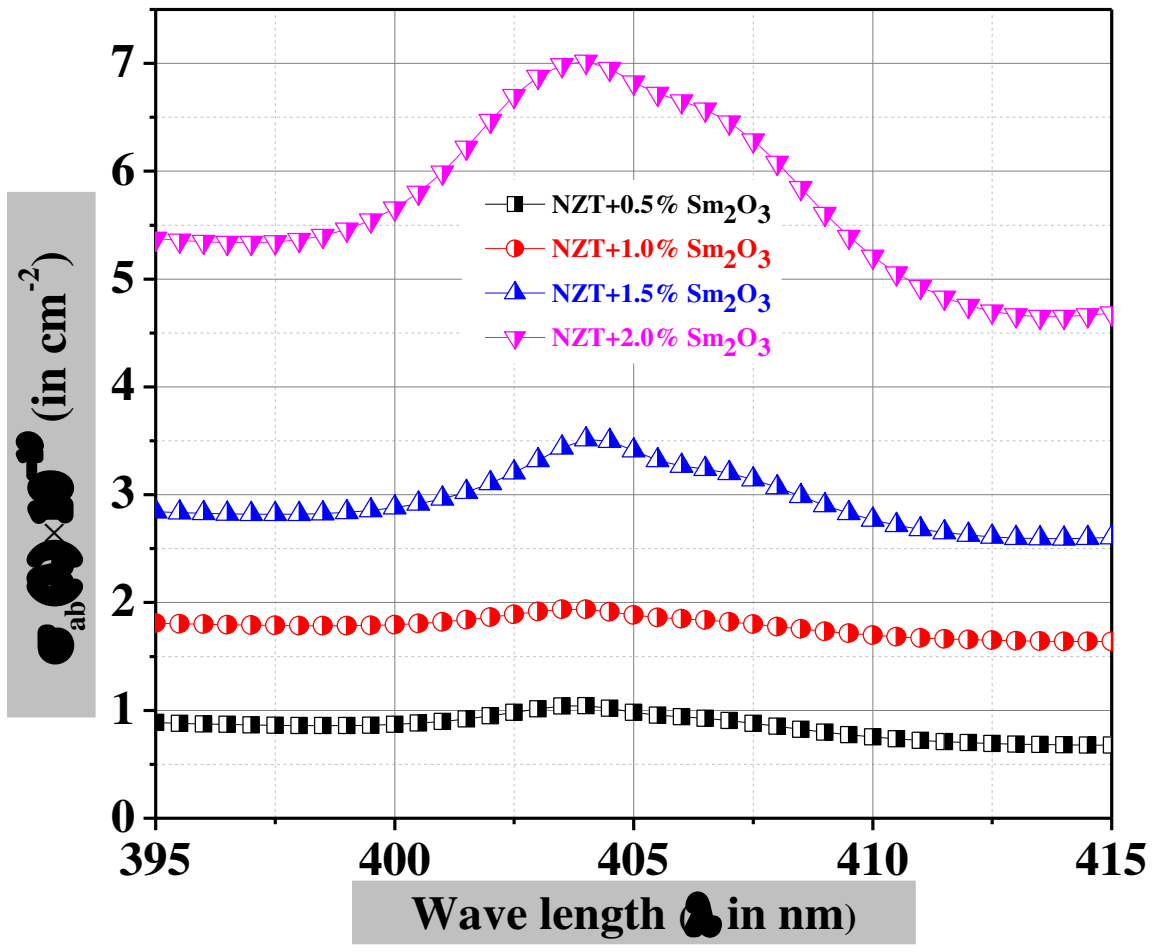

**Figure 11** Wavelength dependence absorption cross-section of $Sm_2O_3$ doped NZT glass materials.

The absorption and emission cross-section are related with the formula

$$\sigma_{emi}(\lambda) = \sigma_{ab}(\lambda) exp\left[\frac{Eg-hv}{K_BT}\right] \ldots \ldots \ldots \quad (9)$$

Where $v$ be the photon frequency, $E_g$ is the free energy needed to excite $Sm^{3+}$ from $^4G_{5/2}$–$^6H_{7/2}$ state at temperature $T$, $h$ is the Planck's constant, and $K_B$ is the Boltzmann constant. Figure 12 shows the emission cross-section of $Sm^{3+}$ doped glasses for the ${}^4G_{5/2}$–${}^6H_{7/2}$ transition. The emission cross-section of $Sm^{3+}$doped NZT glasses is increased due to the higher concentration of $Sm^{3+}$ ions.

The emission cross-section ($\sigma_{emi}$) is the main parameter and its value suggests the rate of energy removal from the lasing material [44]. The higher value of emission cross-section proposes the interesting characteristic for high gain laser applications, low threshold, which are applied to achieve constant wave (CW) laser action.

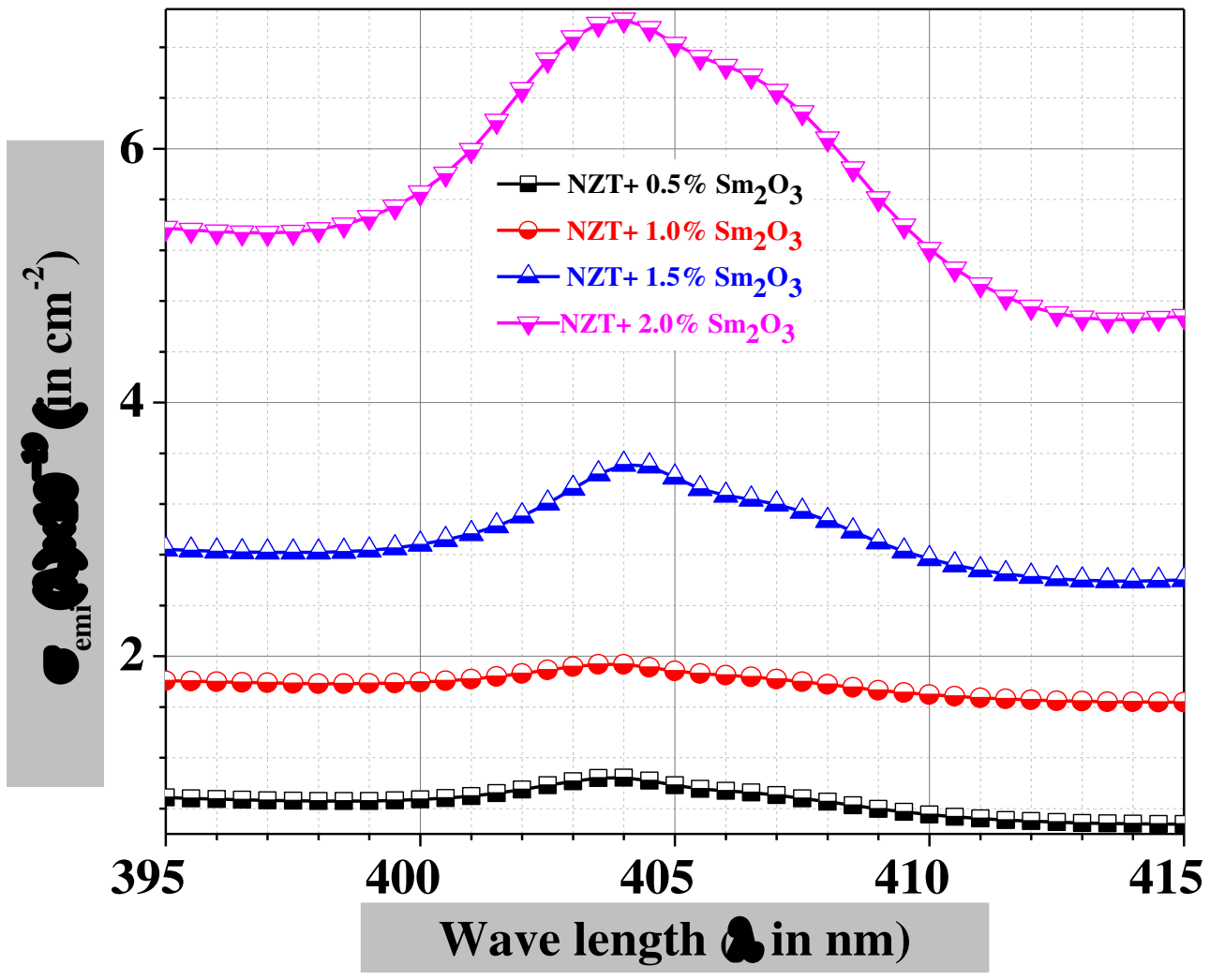

**Figure 12** Wavelength dependence emission cross-section of $Sm_2O_3$ doped NZT glasses.

## 3.4. Dielectric Property

### 3.4.1. Dielectric Constant

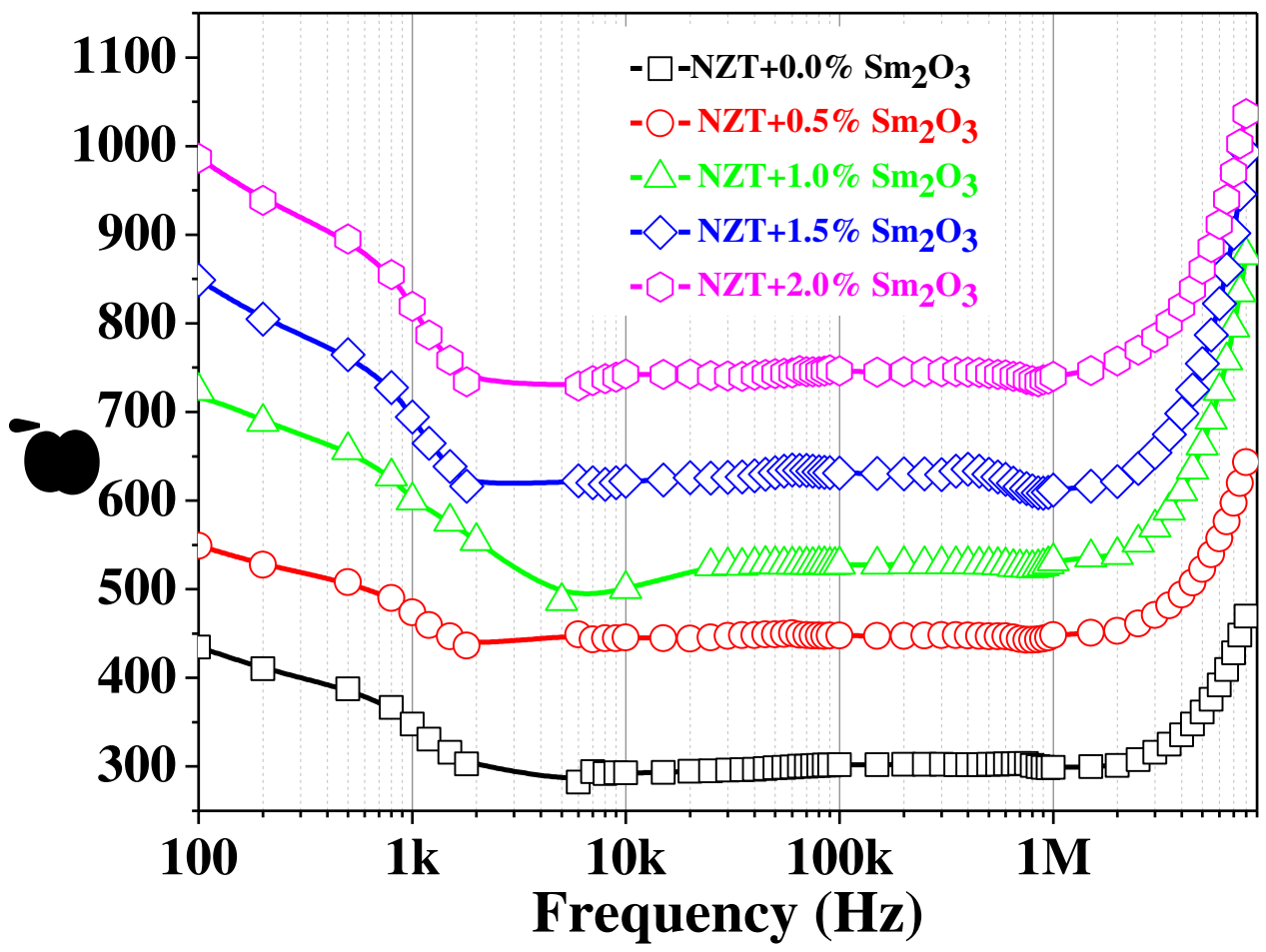

**Figure 13** Dielectric constant (ε′) for $Sm^{3+}$ doped NZT at room temperature.

The values of the dielectric constant are correlated to the electron density and ionic charges. The effect of frequency on dielectric constant for $Sm_2O_3$ doped NZT glass is observed in Fig. 13 for the frequency range100 Hz - 8 MHz at room temperature. It has been observed that the magnitude of ε′ decreases swiftly with increasing frequency of 100 Hz to 2 kHz. This could be attributed to the dielectric dispersion resulting from the delay of the polarization process of the molecules behind the interchanges of the applied electric field [47]. At low frequency, the increase of dielectric constant is also ascribed to the electronic, ionic, dipolar, interfacial polarization and the absence of spontaneous polarization [20, 48]. The magnitude of ε′ of these glasses is almost constant in the frequency region 3 kHz – 2MHz due to the decrease of ionic, space charge and orientation polarization. It has been also described that the pure and doped glasses are not maintaining any variation of dielectric constant with frequency because of the stable molecular orientation and the electron exchange between the ions does not follow high frequency applied field. On the other hand, this may be explained as the applied frequency increases, the ions are not able to respond quickly and it reveals an almost frequency independent behavior may be due to the diminishing number of dipoles which contributes to polarization in the high-frequency region [49].

The dielectric constant increases rapidly beyond the frequency region 2 MHz [50]. It has been observed that the dielectric constant increases from 430 to 980 for $Sm_2O_3$ doping in the lower frequency as well as advanced frequency. This effect may be explained by the charge accumulation due to rare-earth doping. The increasing concentration of doped ions disrupts the tellurite glass system by producing dangling bonds and NBOs. These bonds and NBOs create the motion of charges and thus generate space charge polarization leading to a sharp rise in the dielectric constant [48]. This result is similar to $Eu^{3+}$ doped NZT glass materials [51].

### 3.4.2. DC Conductivity

The variation of electrical conductivity as a function of temperature is shown in fig. 14. The conductivity increases with temperature and also with the doping concentration of $Sm^{3+}$ ions concentration. This variation of conductivity with temperature establishes that the electrical conduction mechanism is Arrhenius type. This Arrhenius mechanism of electrical dc conductivity can be applied to determine the activation energy of the samples using the connection

$$\sigma_{dc} = \sigma_0 exp^{\frac{E_a}{K_B}T} \quad \dots \quad \dots \quad \dots \quad (10)$$

Where $E_a$ is the activation energy, $\sigma_0$ denotes the pre-exponential factor, T is the absolute temperature and $K_B$ is the Boltzmann constant. The assessed activation energies for these samples have been measured from fig. 14. The activation energy reduces (0.63 eV to 0.43 eV) due to the doping of $Sm^{3+}$ ions. The collaboration between the rare-earth ions and structural units of host glass clarified the increment of dc conductivity due to the increasing concentration of rare-earth ions [51]. The technique of conductivity in the glass structure is defined by doping concentration, atomic weight and position of $Sm^{3+}$ ions. The doping concentration of rare-earth ions generates a huge number of NBO in the tellurite glass materials which is also established from FTIR spectra for $Sm^{3+}$ doped NZT glass materials. The higher concentrations of $Sm^{3+}$ ions in the NZT glasses enhance the creation of NBO atoms and hence the conductivity is improved for all the glass materials [52].

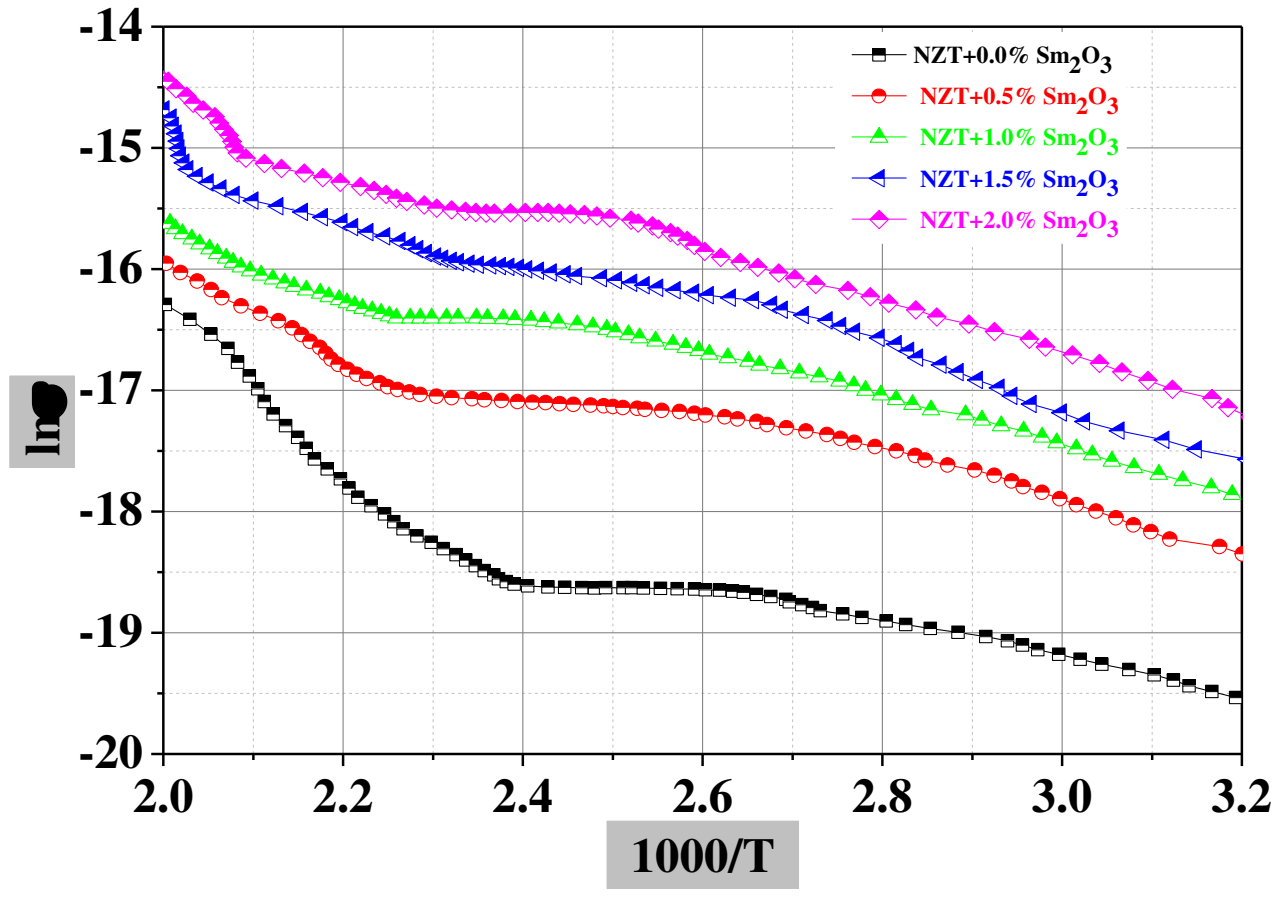

**Figure 14** Arrhenius mechanism of lnσ vs 1000/T.

## 4. Conclusions

Samarium (III) doped NZT glasses have been efficiently prepared by the melt-quenching method. The X-ray diffraction characterization indicates the amorphous nature of the glass materials for the higher doping concentration of $Sm^{3+}$ ions. FTIR spectra reveal the construction of a new glass structure of pure and $Sm_2O_3$ incorporated NZT glasses. The stability of the glass sample is examined by using Hruby's parameter (H) and thermal stability ΔT from the DTA curve. The value of H is 0.57 for highest glass stability in the higher doping of $Sm^{3+}$ ions. The refractive index increases 2.41 to 2.95 with the increase of $Sm^{3+}$ ions. The absorption bands of the glass materials are found corresponding to the transitions $^6H_{5/2}$-$^4P_{3/2}$ at 402 nm, $^6H_{5/2}$-$^4M_{19/2}$ at 418 nm, $^6H_{5/2}$-$^4I_{5/2}$ at 462 nm and $^6H_{5/2}$-$^4I_{11/2}$ at 478 nm. The emissions at 563 nm, $^4G_{5/2}\rightarrow^6H_{5/2}$, at 600 nm, $^4G_{5/2}\rightarrow^6H_{7/2}$, at 645nm, $^4G_{5/2}\rightarrow^6H_{9/2}$ is observed in the photoluminescence spectra. Field strength (F) increases with increasing the $Sm^{3+}$ ions concentration and consequently, the inter-ionic distance ($r_i$) and polaron radius ($r_p$) are observed to decrease. The magnitude of indirect $E_g$ is decreased to the change in the structure of such glasses. The dielectric constant of $Sm_2O_3$ doped NZT glasses has been measured for various frequencies and found as a stable substance within the frequency of 3 kHz-2 MHz. The dc conductivity increases with temperature whereas the activation energy decreases with the increase of ion concentration in the usual manner. So, $Sm_2O_3$ doped NZT glass samples have a large prospect for application in optoelectronic devices.

**Acknowledgement**

The work is partly maintained by DST Govt. of West Bengal research project (Memo No.: 296 (Sanc)/ST/P/S&T/16G- 17/2017) of India. The authors want to thank the CRF, IIT Kharagpur for providing facilities to study DTA and TGA. The authors desire to express thanks to Jadavpur University for providing facilities to study FTIR.

**Figure captions:**

**Figure 1** Picture of $Sm_2O_3$ doped NZT glass sample.

**Figure 2** DTA curves of $Sm_2O_3$ doped NZT glass samples.

**Figure 3** TG with DTG curve of the pure and $Sm_2O_3$ doped NZT glasses.

**Figure 4** XRD pattern of sample pure and $Sm^{3+}$ doped with NZT glass materials.

**Figure 5** FTIR spectroscopy of pure and various $Sm^{3+}$ doped with NZT glass materials.

**Figure 6** Absorption spectroscopy of pure and $Sm^{3+}$ doped with NZT glass materials.

**Figure 7** Tauc's plot for pure and $Sm^{3+}$ doped with NZT glass materials.

**Figure 8** Luminescence spectra of pure and $Sm^{3+}$ doped with NZT glass materials, excited at 402 nm.

**Figure 9** CIE chromaticity color coordinate diagram of $Sm^{3+}$ doped with NZT glass materials, excited at 402 nm.

**Figure 10** Variation of intermolecular distance and field strength with $Sm^{3+}$ ions concentration.

**Figure 11** Wavelength dependence absorption cross-section of $Sm_2O_3$ doped NZT glass materials.

**Figure 12** Wavelength dependence emission cross-section of $Sm_2O_3$ doped NZT glasses. **Figure 13** Dielectric constant (ε′) for $Sm^{3+}$ doped NZT at room temperature.

**Figure 14** Arrhenius mechanism of lnσ vs 1000/T.

**Table captions:**

**Table 1** Thermal parameters calculated from the DTA traces of pure and $Sm_2O_3$ doped NZT glasses.

**Table 2** Band gap energy, refractive index and other physical properties of pure and $Sm^{3+}$ doped NZT glasses.

**Table 1**

| Sample Name | $T_g$ (°C) | $T_c$ (°C) | $T_m$ (°C) | $\Delta T = T_c - T_g$ | $H = (T_c-T_g)/(T_m-T_c)$ |
|---|---|---|---|---|---|
| NZT+0.0% $Sm_2O_3$ | 284 | 328 | 447 | 122 | 0.37 |
| NZT+2.0% $Sm_2O_3$ | 275 | 336 | 443 | 61 | 0.57 |

**Table 2**

| Samples Name | Band gap energy $E_g$ (eV) | Refractive index(n) | Inter Molecular distance ($r_i$) in Å | Polaron Distance ($r_p$) in Å | Field strength(F) $\times 10^{16}$ | Ion density $\times 10^{20}$ |
|---|---|---|---|---|---|---|
| NZT+0.0% $Sm_2O_3$ | 2.95 | 2.41 | -- | -- | -- | -- |
| NZT+0.5% $Sm_2O_3$ | 2.28 | 2.63 | 14.98 | 6.037 | 4.125 | 2.97 |
| NZT+1.0% $Sm_2O_3$ | 2.22 | 2.65 | 13.86 | 5.586 | 4.819 | 3.76 |
| NZT+1.5% $Sm_2O_3$ | 2.12 | 2.69 | 19.51 | 7.862 | 2.432 | 1.35 |
| NZT+2.0% $Sm_2O_3$ | 1.58 | 2.95 | 20.41 | 8.228 | 2.220 | 1.18 |